\documentclass[final,3p,times,twocolumn]{elsarticle}
 \biboptions{comma,sort&compress}
\usepackage{here}
 \usepackage{graphicx}
  \usepackage{epsfig}

\def\nin{\noindent}
\def\beq{\begin{equation}}
\def\eeq{\end{equation}}
\def\bea{\begin{eqnarray}}
\def\eea{\end{eqnarray}}

\def\la{\langle}
\def\ra{\rangle}

\journal{Nuc. Phys. (Proc. Suppl.)}

\begin{document}
\def\be{\begin{eqnarray}}
\def\en{\end{eqnarray}}
\def\la{\langle}
\def\ra{\rangle}
\def\non{\nonumber}
\def\B{{\cal B}}
\def\ov{\overline}
\def\up{\uparrow}
\def\dw{\downarrow}
\def\vp{\varepsilon}
\def\CP{{\it CP}~}
\newcommand{\acp}{\ensuremath{A_{CP}}}

\begin{frontmatter}



\title{The $B$-{\it CP} Puzzles in QCD Factorization}

 \author[label1]{Hai-Yang Cheng\corref{cor1}}
  \address[label1]{Institute of Physics, Academia Sinica, Taipei,
Taiwan, Republic of China}
\cortext[cor1]{Speaker}
\ead{phcheng@phys.sinica.edu.tw}

 \author[label2]{Chun-Khiang Chua}
  \address[label2]{Department of Physics, Chung Yuan Christian University, Chung-Li, Taiwan 320, Republic of China}

\ead{ckchua@phys.cycu.edu.tw}


\begin{abstract}
\noindent
Within the framework of QCD factorization (QCDF), power
corrections due to penguin annihilation can account for the
observed  rates of penguin-dominated two-body decays of $B$ mesons
and direct \CP asymmetries $\acp(K^-\pi^+)$, $\acp(K^{*-}\pi^+)$,
$\acp(K^-\rho^0)$ and $\acp(\pi^+\pi^-)$. However, the predicted
direct {\it CP}-violating effects in QCDF for $B^-\to K^-\pi^0,K^-\eta,\pi^-\eta$ and $\bar B^0\to\pi^0\pi^0$ are wrong in signs
when confronted with experiment. 
We consider two different types of power correction effects in order to resolve the \CP puzzles and rate deficit problems with penguin-dominated two-body decays of $B$ mesons  and color-suppressed tree-dominated $\pi^0\pi^0$ and $\rho^0\pi^0$ modes: penguin annihilation and soft corrections to
the color-suppressed tree amplitude.

\end{abstract}


\end{frontmatter}


\section{Introduction}
\nin
The primary goal and the most important mission of $B$ factories built before millennium is to search for \CP violation in the $B$ meson system. In the past decade, BaBar and Belle have measured direct \CP asymmetries for many charmless hadornic $B$ decays, but only six of them have significance large than 3$\sigma$ and six with significance between $3.0\sigma$ and $1.8\sigma$ (see Table \ref{tab:data}). In the $B_s$ system, CDF has measured $\acp(\bar B_s\to K^+\pi^-)=0.39\pm0.17$ with $S=2.3\sigma$ \cite{CDFcp}.

{\scriptsize
\begin{table}[hbt]
\setlength{\tabcolsep}{0.7pc} \label{tab:data}
 \caption{\scriptsize  Measured \CP asymmetries (in units of \%) and their significance for some of charmless $B$ decays. Data are taken from \cite{HFAG}.}
    {\small
\begin{tabular}{l c c c c}
\hline
 & $K^-\pi^+$ & $\pi^+\pi^-$ & $K^-\eta$ & $\bar K^{*0}\eta$   \\
\hline
$\acp(\%)$ & $-9.8^{+1.2}_{-1.1}$ & $38\pm6$ & $-37\pm9$ & $19\pm5$ \\
$S$ & $8.5\sigma$ & $6.3\sigma$ & $4.1\sigma$ & $3.8\sigma$ \\
\hline
 & $K^-\rho^0$ & $\rho^\pm\pi^\mp$ & $K^{*-}\pi^+$ & $\rho^+K^-$   \\
\hline
$\acp(\%)$ & $37\pm11$ & $-13\pm4$ & $-18\pm7$ & $15\pm6$ \\
$S$ & $3.4\sigma$ & $3.3\sigma$ & $2.6\sigma$ & $2.5\sigma$ \\
\hline
 & $K^-\pi^0$ & $\pi^-\eta$ & $\pi^0\pi^0$ & $\rho^-\pi^+$   \\
\hline
$\acp(\%)$ & $5.0\pm2.5$ & $-13\pm7$ & $43^{+25}_{-24}$ & $11\pm6$ \\
$S$ & $2.0\sigma$ & $1.9\sigma$ & $1.8\sigma$ & $1.8\sigma$ \\
\hline
\end{tabular}
}
\label{tab:param}
\end{table}
}
\nin

The quantity
$\Delta
A_{K\pi}\equiv\acp(K^-\pi^0)-\acp(K^-\pi^+)$ measures the \CP asymmetry difference of $B^-\to K^-\pi^0$ and $\bar B^0\to K^-\pi^+$. Its magnitude is $(14.8\pm2.8)\%$. Naively, it is expected that
$\acp(K^-\pi^0)\approx \acp(K^-\pi^+)$, whereas experimentally
$\Delta A_{K\pi}$ differs from zero by 5.3$\sigma$ effect. This is the so-called $K\pi$ \CP puzzle.

Generally,
the physics behind nonleptonic $B$ decays is extremely complicated. Nevertheless, it is greatly simplified in the heavy quark limit $m_b\to\infty$ as the decay amplitude becomes factorizable and can be expressed in terms of decay constants and form factors. However, this simply approach encounters several major difficulties:
(i) the predicted branching fractions for penguin-dominated $\bar B\to PP,VP,VV$ decays are systematically below the measurements \cite{BN} and the rates for color-suppressed tree-dominated decays $\bar B^0\to\pi^0\pi^0,\rho^0\pi^0$ are too small,   (ii) direct {\it CP}-violating asymmetries for $\bar B^0\to K^-\pi^+$, $\bar B^0\to K^{*-}\pi^+$, $B^-\to K^-\rho^0$, $\bar B\to \pi^+\pi^-$ and $\bar B_s^0\to K^+\pi^-$ disagree with experiment in signs, and (iii) the transverse polarization fraction in penguin-dominated charmless $B\to VV$ decays is predicted to be very small, while experimentally it is comparable to the longitudinal polarization one. All these indicate the necessity of going beyond zeroth $1/m_b$ power expansion.

Let us first consider power corrections to the QCD penguin amplitude of the $\bar B\to PP$ decay which has the generic expression
\be \label{eq:P}
P&=& P_{\rm SD}+P_{\rm LD} \non \\
&=& A_{PP}[\lambda_u(a_4^u+r_\chi^P a_6^u)+\lambda_c(a_4^c+r_\chi^P a_6^c)]\non \\
&& +1/m_b~{\rm corrections},
\en
where $\lambda_p^{(q)}=V_{pb}V_{pq}^*$ with $q=s,d$, $a_{4,6}$ are the effective Wilson coefficients and  $r^P_\chi$ is a chiral factor of order unity. Possible power corrections to penguin amplitudes include long-distance charming penguins, final-state interactions and  penguin annihilation characterized by the parameters $\beta_3^{u,c}$. In the so-called ``S4" scenario of QCDF \cite{BN}, power corrections to the penguin annihilation topology characterized by $\lambda_u\beta_3^u+\lambda_c\beta_3^c$ are added to Eq. (\ref{eq:P}). By adjusting the magnitude and phase of $\beta_3$ in this scenario, all the above-mentioned discrepancies except for the rate deficit problem with the decays $\bar B^0\to\pi^0\pi^0,\rho^0\pi^0$ can be resolved.

\section{New \CP puzzles}
\nin

However, a scrutiny of the QCDF predictions reveals more puzzles with respect to direct \CP violation. While the signs of \CP asymmetries in $K^-\pi^+,K^-\rho^0$ modes are flipped to the right ones in the presence of power corrections from penguin annihilation,
the signs of $\acp$ in $B^-\to K^-\pi^0,~K^-\eta,~\pi^-\eta$ and $\bar B^0\to\pi^0\pi^0,~\bar K^{*0}\eta$ will also get reversed in such a way that they disagree with experiment. In other words, in the heavy quark limit the \CP asymmetries of these five  modes  have the right signs when compared with experiment.

The aforementioned direct \CP puzzles indicate that it is necessary to consider subleading power corrections other than penguin annihilation. For example, the large power corrections due to $P'$ cannot explain the $\Delta A_{K\pi}$ puzzle as they contribute equally to both $B^-\to K^-\pi^0$ and $\bar B^0\to K^-\pi^+$. The additional power correction should have little effect on the decay rates of penguin-dominated decays but will manifest in the measurement of direct \CP asymmetries.
Note that all the "problematic" modes receive a contribution from $c^{(')}=C^{(')}+P_{\rm EW}^{(')}$. Since
$A(B^-\to K^-\pi^0)\propto t'+c'+p'$ and $A(\bar B^0\to K^-\pi^+)\propto t'+p'$ with $t'=T'+P'^c_{\rm EW}$ and
$p'=P'-{1\over 3}P'^c_{\rm EW}+P'_A$, we can consider this puzzle resolved, provided that $c'/t'$ is of order
$1.3\sim 1.4$ with a large negative phase ($|c'/t'|\sim 0.9$ in the standard short-distance effective Hamiltonian approach). There are several  possibilities for a
large complex $c'$: either a large complex $C'$ or a large complex electroweak penguin
$P'_{\rm EW}$ or a combination of them. Various scenarios
for accommodating large $C'$ \cite{Li05,Kim,Li09,Baek09} or $P'_{\rm EW}$ \cite{LargeEWP} have
been proposed. To get a large complex $C'$, one can resort to spectator scattering or final-state interactions. However, the general consensus for a large complex $P'_{\rm EW}$ is that one needs New Physics beyond the Standard Model because it is well known that $P'_{\rm EW}$ is essentially real in the SM as it does not carry a nontrivial strong phase \cite{NR}.
In principle, one cannot discriminate between these two  possibilities in penguin-dominated decays as it is always the combination $c'=C'+P'_{\rm EW}$ that enters into the decay  amplitude except for the decays involving
$\eta$ and/or $\eta'$ in the final state where
%
%
both $c'$ and $P'_{\rm EW}$ present in the
amplitudes~\cite{Chiang}.  Nevertheless, these two scenarios will lead to very distinct predictions for tree-dominated decays where $P_{\rm EW}\ll C$. (In penguin-dominated decays, $P'_{\rm EW}$ is comparable to $C'$ due to the fact that $\lambda_c^{(s)}\gg\lambda_u^{(s)}$.) The decay rates of $\bar B^0\to \pi^0\pi^0,\rho^0\pi^0$ will be substantially enhanced for a large $C$ but remain intact for a large $P_{\rm EW}$. Since $P_{\rm EW}\ll C$ in tree-dominated
channels, \CP puzzles with $\pi^-\eta$ and $\pi^0\pi^0$ cannot be
resolved with a large $P_{\rm EW}$. Therefore, it is most likely
that the  color-suppressed tree amplitude is large and complex.
In other words, the $B\to K\pi$ \CP puzzle can be resolved without invoking New Physics.

It should be remarked that the aforementioned $B$-{\it CP} puzzles with the $K^-\pi^0,~K^-\eta,~\pi^-\eta,~\bar K^{*0}\eta,~\pi^0\pi^0$ modes also occur in the approach of soft-collinear effective theory (SCET) \cite{SCET} where the penguin annihilation effect in QCDF is replaced by the long-distance charming penguins. The $B$-{\it CP} puzzles mentioned here
are relevant to QCDF and may not occur in other approaches such as
pQCD \cite{pQCD}
owing to a different treatment of endpoint divergence in penguin annihilation diagrams (see Table \ref{tab:Theory} below)

In this work we shall consider
the possibility of having a large color-suppressed tree amplitude with a sizable strong phase relative to the color-allowed tree amplitude \cite{CCcp}
\be
C=[\lambda_u a_2^u]_{\rm SD}+[\lambda_u a_2^u]_{\rm LD}+{\rm FSIs}+\cdots.
\en
As will be discussed below, the long-distance contribution to $a_2$ can come from the twist-3 effects in spectator rescattering, while an example of final-state rescattering contribution to $C$ will be illustrated below.

\section{Power corrections to $a_2$}
\nin
Following \cite{CCcp}, power corrections to the color-suppressed topology are parametrized as \be \label{eq:a2}
a_2 \to a_2(1+\rho_C e^{i\phi_C}),
\en
with the unknown parameters $\rho_C$ and $\phi_C$ to be inferred from experiment. We shall use \cite{CCcp}
\be
 \rho_C\approx 1.3\,,~0.8\,,~0, \qquad
 \phi_C\approx -70^\circ\,,-80^\circ\,,~0,
\en
for $\bar B\to PP,VP,VV$ decays, respectively.
This pattern that soft power corrections to $a_2$ are large for $PP$ modes, moderate for $VP$ ones and very small for $VV$ cases is consistent with the observation made in \cite{Kagan} that soft power correction dominance is much larger for $PP$ than $VP$ and $VV$ final states.
It has been argued that this has to do with the special nature of the pion which is a $q\bar q$ bound state on the one hand and a nearly massless Nambu-Goldstone boson on the other hand \cite{Kagan}.

What is the origin of power corrections to $a_2$ ? There are two possible sources: hard spectator interactions and final-state interactions. From Eq. (\ref{eq:ai}) we have the expression
\be \label{eq:ai}
  a_2(M_1M_2) &=&
 c_2+{c_1\over N_c}
  + {c_1\over N_c}\,{C_F\alpha_s\over
 4\pi}\Big[V_2(M_2) \\
 &+&{4\pi^2\over N_c}H_2(M_1M_2)\Big]+a_2(M_1M_2)_{\rm LD}, \non
\en
for $a_2$.
The hard spectator term $H_2(M_1 M_2)$ reads
\begin{eqnarray}\label{eq:hardspec}
 && H_2(M_1 M_2)= {if_B f_{M_1} f_{M_2} \over X^{(\overline{B} M_1,
  M_2)}}\,{m_B\over\lambda_B} \int^1_0 d x d y \non \\
 &&\times\Bigg( \frac{\Phi_{M_1}(x) \Phi_{M_2}(y)}{\bar x\bar y}
  + r_\chi^{M_1}
  \frac{\Phi_{m_1} (x) \Phi_{M_2}(y)}{\bar x y}\Bigg),
 \hspace{0.5cm}
 \end{eqnarray}
where $X^{(\overline{B} M_1,
  M_2)}$ is the factorizable amplitude for $\ov B\to M_1M_2$, $\bar x=1-x$.
Power corrections from the twist-3 amplitude $\Phi_m$ are divergent and can be parameterized as
 \be \label{eq:XH}
 X_H\equiv \int^1_0{dy\over y}={\rm ln}{m_B\over
 \Lambda_h}(1+\rho_H e^{i\phi_H}).
 \end{eqnarray}
Since $c_1\sim {\cal O}(1)$ and $c_9\sim {\cal O}(-1.3)$ in units of $\alpha_{em}$, it is clear that hard spectator contributions to $a_i$ are usually very small except for $a_2$ and $a_{10}$. Indeed, there is a huge cancelation between the vertex and naive factorizable terms so that the real part of $a_2$ is governed
by spectator interactions, while its imaginary part comes mainly from the vertex corrections \cite{BBNS2}. The value of $a_2(K\pi)\approx 0.51\,e^{-i58^\circ}$ needed to solve the $B\to K\pi$ \CP puzzle  corresponds to $\rho_H\approx 4.9$ and $\phi_H\approx -77^\circ$. Therefore, there is no reason to restrict $\rho_H$ to the range $0\leq \rho_H\leq 1$. A sizable color-suppressed tree amplitude also can be induced via color-allowed decay $B^-\to K^-\eta'$ followed by the rescattering of $K^-\eta'$ into $K^-\pi^0$.
Recall that among the 2-body $B$ decays, $B\to K\eta'$ has the largest branching fraction, of order $70\times 10^{-6}$.
This final-state rescattering has the same topology as the color-suppressed tree diagram \cite{CCSfsi}. One of us (CKC) has studied the FSI effects through residual rescattering among $PP$ states and resolved the $B$-$CP$ puzzles \cite{Chua}.

\section{$K\pi$ \CP puzzle}
The \CP asymmetry of $\bar B^0\to K^-\pi^+$ can be expressed as
\be \label{eq:acpKpi}
\acp(\bar B^0\to K^-\pi^+)\,R_{\rm FM}=-2\sin\gamma\,{\rm Im}\,r_{\rm FM},
\en
with
\be \label{eq:rFM}
R_{\rm FM} &\equiv& {\Gamma(\bar B^0\to K^-\pi^+)\over \Gamma(B^-\to \bar K^0\pi^-)}=1-2\cos\gamma\,{\rm Re}\, r_{\rm FM} \non \\
&& \qquad\qquad\qquad\quad +|r_{\rm FM}|^2, \non \\
r_{\rm FM} &=& \left|{\lambda_u^{(s)}\over \lambda_c^{(s)}}\right|\,{\alpha_1(\pi \bar K)\over -\alpha_4^c(\pi\bar K)-\beta_3^c(\pi\bar K)},
\en
In the absence of penguin annihilation, direct \CP violation of $\bar B^0\to K^-\pi^+$ is positive as Im\,$\alpha_4^c\approx 0.013$. When the power correction to penguin annihilation is turned on, we have Im$(\alpha_4^c+\beta^c_3)\approx -0.039$ and hence a negative $\acp(K^-\pi^+)$. This also explains why \CP asymmetries of penguin-dominated decays in the QCDF framework will often reverse their signs in the presence of penguin annihilation.

The decay amplitude of $B^-\to K^-\pi^0$ is
\be
\sqrt{2}A(B^-\to K^-\pi^0) &=& A_{\pi\bar K}(\delta_u \alpha_1+\alpha_4^p+\beta_3^p) \\
&+& A_{\bar K\pi}(\delta_{pu}\alpha_2+{3\over 2}\alpha^p_{\rm 3,EW}). \non
\en
If the color-suppressed tree and electroweak penguin amplitudes are negligible, it is obvious that the amplitude of $K^-\pi^0$ will be the same as that of $K^-\pi^+$ except for a trivial factor of $1/\sqrt{2}$.
\be \label{eq:acpKpi}
\acp(\bar B^0\to K^-\pi^+)\,R_{\rm FM}&=&-2\sin\gamma\,{\rm Im}\,r_{\rm FM} \non \\
&& -2\sin\gamma{\rm Im}r_C,
\en
where
\be
r_C=\left|{\lambda_u^{(s)}\over \lambda_c^{(s)}}\right|\,{f_\pi F_0^{BK}(0)\over f_K F_0^{B\pi}(0)}{\alpha_2(\pi \bar K)\over -\alpha_4^c(\pi\bar K)-\beta_3^c(\pi\bar K)}.
\en
The imaginary part of $r_C$ is rather small because of the cancelation of the phases between $\alpha_2$ and $\alpha_4^c+\beta_3^c$. When soft corrections to $a_2$ are included, we have $r_C\approx 0.078-0.063i$\,. It follows that $\Delta A_{K\pi}$ will become of order 0.13\,.

The QCDF predictions of direct \CP asymmetries are summarized in Table \ref{tab:Theory}. The pQCD results are also listed for comparison. In the pQCD approach, the predictions for some of the $VP$ modes, e.g. $\acp(K^{*-}\pi^+)$, $\acp(\rho^0K^-)$ and $\acp(\rho^+ K^-)$ are very large, above 50\%. This is because QCD penguin contributions in these modes are small, and direct \CP violation arises from the interference between tree and annihilation diagrams. The strong phase comes mainly from the annihilation diagram in this approach.

{\scriptsize
\begin{table}[hbt]
\setlength{\tabcolsep}{0.6pc}
 \caption{\scriptsize   The predicted direct \CP asymmetries (in \%) in QCD factorization and pQCD taken from \cite{CCBud}. Experimental measurements are from \cite{HFAG}.} \label{tab:Theory}
    {\small
\begin{tabular}{l c c c c}
\hline
 & $K^-\pi^+$ & $\pi^+\pi^-$ & $K^-\eta$ & $\bar K^{*0}\eta$   \\
\hline
Expt & $-9.8^{+1.2}_{-1.1}$ & $38\pm6$ & $-37\pm9$ & $19\pm5$ \\
QCDF & $-7.4^{+4.6}_{-5.0}$ & $17.0^{+4.5}_{-8.8}$ & $-11.2^{+17.4}_{-24.3}$ & $3.5^{+2.7}_{-2.4}$ \\
pQCD & $-10^{+7}_{-8}$ & $18^{+20}_{-12}$ & $-11.7^{+~8.4}_{-10.5}$ & $4.60^{+1.16}_{-1.32}$ \\
\hline
 & $K^-\rho^0$ & $\rho^\pm\pi^\mp$ & $K^{*-}\pi^+$ & $\rho^+K^-$   \\
\hline
Expt & $37\pm11$ & $-13\pm4$ & $-18\pm7$ & $15\pm6$ \\
QCDF & $45.4^{+36.1}_{-30.2}$ & $-11^{+7}_{-5}$ & $-12.1^{+12.6}_{-16.0}$ & $31.9^{+22.7}_{-16.8}$ \\
pQCD & $71^{+25}_{-35}$ & -- & $-60^{+32}_{-19}$ & $64^{+24}_{-30}$ \\
\hline
 & $K^-\pi^0$ & $\pi^-\eta$ & $\pi^0\pi^0$ & $\rho^-\pi^+$   \\
\hline
Expt & $5.0\pm2.5$ & $-13\pm7$ & $43^{+25}_{-24}$ & $11\pm6$ \\
QCDF & $4.9^{+5.9}_{-5.8}$ & $-5.0^{+~8.7}_{-10.8}$ & $57.2^{+33.7}_{-40.4}$ & $4.4^{+5.8}_{-6.8}$ \\
pQCD & $-1^{+3}_{-6}$ & $-37^{+9}_{-7}$ & $63^{+35}_{-34}$ & -- \\
\hline
\end{tabular}
}
\end{table}
}

\section{Mixing-induced \CP violation}
Possible New Physics beyond the Standard Model is being
intensively searched via the measurements of time-dependent \CP
asymmetries in neutral $B$ meson decays into final \CP eigenstates
defined by
 \be
 {\Gamma(\ov B(t)\to f)-\Gamma(B(t)\to f)\over
 \Gamma(\ov B(t)\to f)+\Gamma(B(t)\to
 f)} &=& S_f\sin(\Delta mt)  \\
 &-& C_f\cos(\Delta mt), \non
 \en
where $\Delta m$ is the mass difference of the two neutral $B$
eigenstates, $S_f$ monitors mixing-induced \CP asymmetry and
$A_f$ measures direct \CP violation (note that $C_f=-\acp$).
Since the theoretical calculation of $S_f$ depends on the input of the angle $\beta$ or $\sin 2\beta$, it is more sensible to consider the difference
\be
\Delta S_f\equiv -\eta_f S_f-\sin 2\beta
\en
for penguin-dominated decays.

Predictions and the data of $\Delta S_f$ are listed in Table \ref{tab:DeltaSPP}.
The decay modes $\eta'K_S$ and $\phi K_S$ appear theoretically very clean in QCDF;  for these modes
the central value of $\Delta S_f$ as well as the uncertainties are rather small. By sharp contrast, the theoretical errors in pQCD predictions for both $S_{\eta'K_S}$ and $S_{\eta K_S}$  arising from uncertainties in the CKM angles $\alpha$ and $\gamma$ are very large \cite{XiaoKeta}. This issue should be resolved.

{\scriptsize
\begin{table}[!htbp]
\setlength{\tabcolsep}{0.1pc}
\caption{Mixing-induced \CP\ violation $\Delta S_f$  in $B\to PP,VP$ decays predicted in QCDF and pQCD.
 Experimental results are from BaBar (first entry) and Belle (second entry) \cite{HFAG}. }\label{tab:DeltaSPP}
 {\small
\begin{tabular}{l c c c c} \hline
Mode & QCDF & pQCD & Expt & Average  \\ \hline
 $\eta' K_S$
       & $0.00^{+0.01}_{-0.01}$ & $-0.06^{+0.50}_{-0.91}$
       & $\begin{array}{c}-0.10\pm0.08\\ -0.03\pm0.11\end{array}$
       & $-0.08\pm0.07$
       \\
 $\eta K_S$
       & $0.12^{+0.09}_{-0.08}$ & $-0.07^{+0.50}_{-0.92}$
       \\
 $\pi^0K_S$
       & $0.12^{+0.07}_{-0.06}$ & $0.06^{+0.02}_{-0.03}$
       & $\begin{array}{c}-0.12\pm0.20\\ 0.00\pm0.32\end{array}$
       & $-0.10\pm0.17$
       \\
 $\phi K_S$
       & $0.022^{+0.004}_{-0.002}$ & $0.02\pm0.01$
       & $\begin{array}{c}-0.41\pm0.26\\ 0.23^{+0.09}_{-0.19}\end{array}$
       & $-0.11^{+0.16}_{-0.18}$
       \\
 $\omega K_S$
       & $0.17^{+0.06}_{-0.08}$ & $0.15^{+0.03}_{-0.07}$
       & $\begin{array}{c}-0.12^{+0.26}_{-0.29}\\ -0.56\pm0.47\end{array}$
       & $-0.22\pm0.24$
       \\
 $\rho^0K_S$
       & $-0.17^{+0.09}_{-0.18}$ & $-0.19^{+0.10}_{-0.06}$
       & $\begin{array}{c}-0.32^{+0.27}_{-0.31}\\ -0.03^{+0.23}_{-0.28}\end{array}$
       & $-0.13^{+0.18}_{-0.21}$
       \\
       \hline
\end{tabular}
}
\end{table}
}

\section*{Acknowledgements}
\nin
This research was supported in part by the National Science Council of Taiwan, R.~O.~C. and in part by the NCTS.


\end{document}